\documentclass[12pt]{iopart}
\usepackage{iopams}  
\usepackage{graphicx}
\usepackage{epstopdf}
\begin{document}


\title[The graphene-ferromagnet interface]{Electronic and magnetic properties of the graphene-ferromagnet interface}

\author{Yu. S. Dedkov$^1$ and M. Fonin$^2$}
\address{$^1$Fritz-Haber-Institut der Max-Planck-Gesellschaft, Berlin, Germany}
\address{$^2$Fachbereich Physik, Universit\"at Konstanz, Germany}
\ead{dedkov@fhi-berlin.mpg.de}

\begin{abstract}
The article presents the work on the investigation of the surface structure as well as electronic and magnetic properties of graphene layer on a lattice matched surface of a ferromagnetic material, Ni(111). Scanning tunneling microscopy imaging shows that perfectly ordered epitaxial graphene layers can be prepared by elevated temperature decomposition of hydrocarbons, with domains larger than the terraces of the underlying Ni(111). In some exceptional cases graphene films do not show rotational alignment with the metal surface leading to moir\'e structures with small periodicities. We give a detailed analysis of the crystallographic structure of graphene with respect to the Ni(111) surface based both on experimental results and recent theoretical studies. X-ray absorption spectroscopy investigations of empty valence band states demonstrate the existence of interface states which originate from the strong hybridization between the graphene $\pi$ and Ni $3d$ valence-band states with the partial charge transfer of the spin-polarized electrons to the graphene $\pi^*$ unoccupied states. The latter leads to the appearance of an induced magnetic moment of carbon atoms in the graphene layer which is unambiguously confirmed by both x-ray magnetic circular dichroism and spin-resolved photoemission. Further angle-resolved photoemission investigations indicate a strong interaction between graphene and Ni(111) showing considerable modification of the valence-band states of Ni and graphene due to a strong hybridization. The detailed analysis of the Fermi surface of the graphene/Ni(111) system show very good agreement between experimental and calculated two-dimensional maps of the electronic states around the Fermi level confirming the main predictions of the theory. We analyze our spectroscopic results relying on the currently available band structure calculations for the graphene/Ni(111) system and discuss the perspectives of the realization of graphene/ferromagnet-based devices.
\end{abstract}


\maketitle

\section{Introduction}\label{introduction}

Graphene is a two-dimensional sheet of carbon atoms arranged in a honeycomb lattice with two atoms in the unit cell~\cite{Geim:2007a,CastroNeto:2009,Geim:2009}. The $sp^2$ hybridization between one $s$ orbital and two $p$ orbitals leads to a trigonal planar structure with a formation of a $\sigma$ bonds between carbon atoms that are separated by $1.42$\,\AA. These bands have a filled shell and, hence, form a deep valence band. The half-filled $p_z$ orbitals, which are perpendicular to the planar structure, form the bonding ($\pi$) and antibonding ($\pi^*$) bands. The $\pi$ and $\pi^*$ bands touch in a single point exactly at the Fermi energy ($E_F$) at the corner of the hexagonal graphene's Brillouin zone. Close to this so-called Dirac point ($E_D$) the bands display a linear dispersion and form perfect Dirac cones~\cite{CastroNeto:2009}. Thus, undoped graphene is a semimetal (``zero-gap semiconductor''). The linear dispersion of the bands mimics the physics of quasiparticles with zero mass, so-called Dirac fermions~\cite{Geim:2007a,CastroNeto:2009,Geim:2009}.

The exceptional transport properties of graphene~\cite{Geim:2007a} make it a promising material for applications in microelectronics~\cite{Novoselov:2005,Morozov:2008} and sensing~\cite{Schedin:2007}. This has recently led to a revival of interest in graphene on transition metal surfaces~\cite{Gamo:1997,Dedkov:2008a,Dedkov:2008b,Martoccia:2008,Sutter:2009,Wang:2008,Ndiaye:2006,Coraux:2008,Coraux:2009,Hu:1987,Land:1992,Sasaki:2000}, as large area epitaxial graphene layers of exceptional quality can be grown, which might be an alternative to micromechanical cleavage for producing macroscopic graphene films. The first mass-production of high quality graphene layers via chemical-vapour deposition (CVD) method on polycrystalline Ni surface and its transferring to arbitrary substrate was demonstrated in the beginning of 2009~\cite{Kim:2009a}. The transferred graphene films showed very low sheet resistance of 280\,$\Omega$ per square, with 80\% optical transparency, high electron mobility of 3700\,cm$^2$V$^{-1}$s$^{-1}$, and the half-integer quantum Hall effect at low temperatures indicating that the quality of graphene grown by CVD is as high as mechanically cleaved graphene. Further modifications of this method allow the preparation of predominantly monolayer graphene films with a size of more than 30 inch which can be further transferred on a polymer film for the fabrication of transparent electrodes~\cite{Bae:2010}.

The electronic interaction of graphene with a metal is both of fundamental and technological interest in view of possible device applications. Recent theoretical calculations by V. M. Karpan and co-workers~\cite{Karpan:2007,Karpan:2008} for graphene/metal interfaces imply the possibility of an ideal spin-filtering in the current-perpendicular-to-the-plane configuration (CPP) for the ferromagnet/graphene/ferromagnet sandwich-like structures. The close-packed surfaces of Co and Ni were considered as ferromagnetic (FM) electrodes which perfectly coincide with graphene from the crystallographic point of view (Fig.\,1). The spin-filtering effect originates form the unique overlapping of the electronic structures of the graphene monolayer and close-packed surfaces of ferromagnetic Ni and Co. As discussed earlier, graphene is a semimetal with electronic density in the vicinity of $E_F$ at corners ($K$ points)  of the hexagonal Brillouin zone of graphene (Fig.\,2). If the Fermi surface projections of ferromagnetic metals, $fcc$ Ni or Co, on the (111) close-packed plane are considered, then in both cases graphene has only minority electron density around the $K$ points of the surface Brillouin zone. In this case, it is expected that in the absence of any additional factors that lower the symmetry of the system, the preferential transport of only minority electrons and perfect spin-filtering will appear in a FM/graphene/FM stack~\cite{Karpan:2007,Karpan:2008}. The interaction between graphene and ferromagnetic material will however change the electronic properties of the interface partially quenching the spin-filtering effect in the sandwich-like structure, but a sizable effect can still be detected by choosing the proper combination of FM materials~\cite{Yazyev:2009} and this effect is predicted to increase strongly when multilayer graphene is used~\cite{Karpan:2007,Karpan:2008}. 

Besides spin-filtering, graphene might be the best material for the realization of spintronic devices. Such systems usually require the effective injection of the spin-polarized electrons in the conductive channel which can be made from graphene~\cite{Tombros:2007}. However, prior to being able to implement graphene/ferromagnet systems in any kind of spintronic unit, a study of the electronic, magnetic, and interfacial properties has to be performed.

\begin{figure}
\center
\includegraphics[scale=0.275]{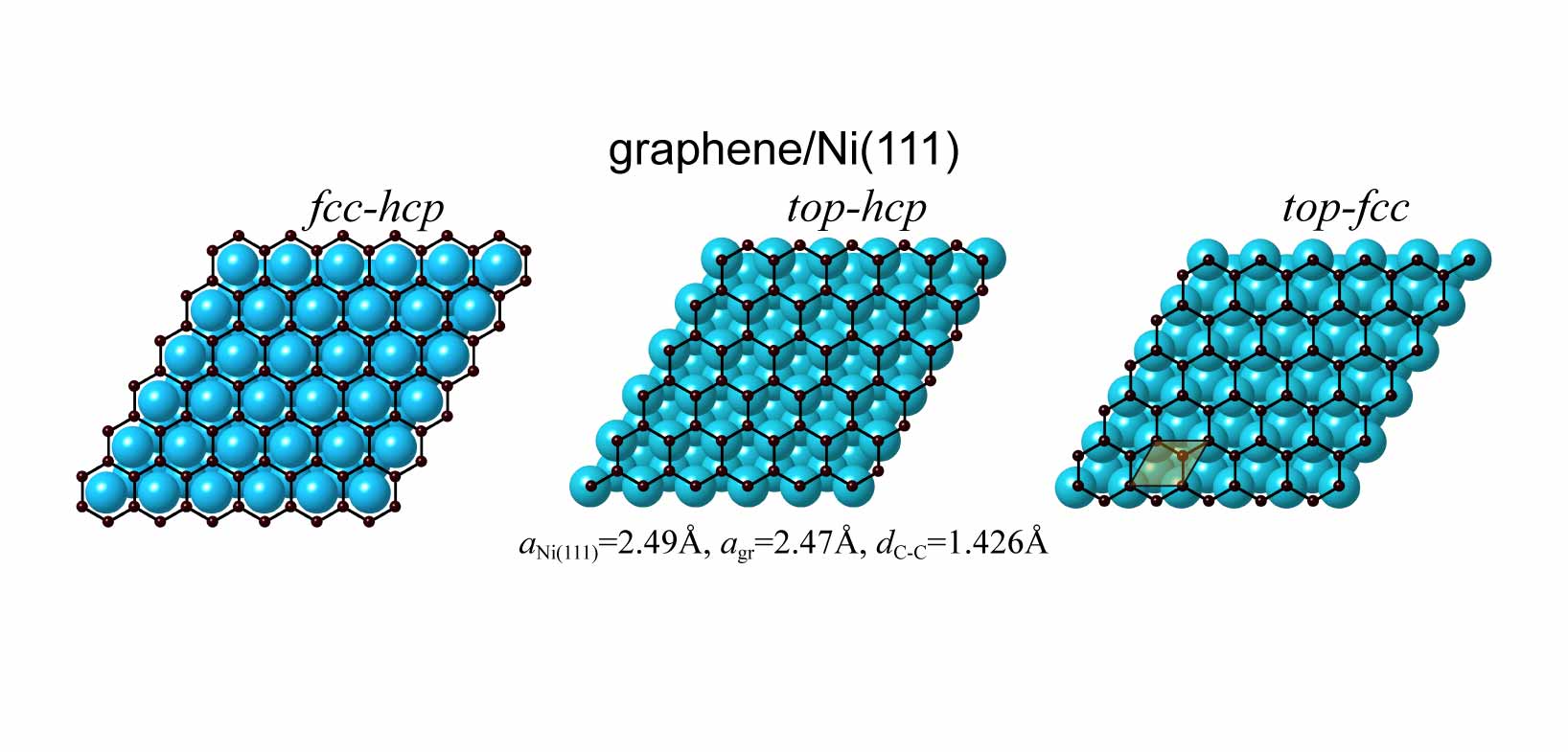}
\caption{Top view of a ball model showing three possible arrangements of the graphene layer on top of the close-packed Ni(111) surface. All structures have three-fold symmetry. Left-hand panel: carbon atoms are located above Ni atoms in the 2nd and 3rd Ni layers - $fcc-hcp$ structure; middle panel: carbon atoms are above Ni atoms in the 1st and 2nd layers - $top-hcp$ structure; right-hand panel: carbon atoms are placed above Ni atoms in the 1st and 3rd layers - $top-fcc$ structure. Carbon atoms are small black spheres; first, second and third layer Ni atoms are large blue spheres.}
\end{figure}

\begin{figure}
\center
\includegraphics[scale=0.6]{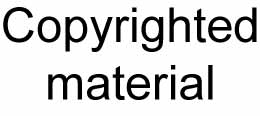}
\caption{Left-hand panel: Fermi surface projections onto the (111) close-packed plane for (a,b) $fcc$ Co (majority and minority spins), (c,d) $fcc$ Ni (majority and minority spins), and (e) $fcc$ Cu. For graphene (f), the constant-energy surface is centered in the $K$ point of the two-dimensional interface Brillouin zone. Right-hand panel: the results of band structure calculations (majority and minority spins) for (g) $fcc-hcp$ (BC) and (h,i) $top-fcc$ (AC) configurations of carbon atoms on Ni(111). The weight of the carbon $\pi_z$ character is shown by black lines where its thickness reflects corresponding orbital weight. Reprinted figures with permission from V. M. Karpan \textit{et al.}, Phys. Rev. B \textbf{78}, 195419 (2008).  Copyright (2008) by the American Physical Society.}
\end{figure}

In the present paper the crystallographic structure, morphology, electronic and magnetic properties of a graphene/ferromagnet interface are considered for the case of the Ni(111) close-packed surface. STM investigation shows that perfectly ordered epitaxial graphene layers can be prepared on Ni(111). X-ray absorption spectroscopy (XAS) studies of graphene/Ni(111) reveal the existence of interface states which originate from the strong hybridization of the graphene $\pi$ and Ni $3d$ valence-band states with the partial charge transfer of the spin polarized electrons onto the graphene $\pi^*$ unoccupied states. This leads to the appearance of the induced magnetic moment of $\mu=0.05-0.1\mu_B$ on the carbon atoms in the graphene layer that is confirmed by both x-ray magnetic circular dichroism (XMCD) and spin-resolved photoemission (PES). Angle-resolved photoemission (ARPES) data confirm the strong interaction between graphene and Ni(111) showing considerable modification of the valence-band states of Ni and graphene due to hybridization. The three-dimensional (3D) mapping of electronic states gives full information about the band structure of the graphene/Ni(111) system. Detailed analysis of the Fermi surface of the graphene/Ni(111) system indicates very good agreement between experimental and calculated 2D pictures of electronic states in the valence band.

\section{Experimental details}\label{experimental}

The present studies of the graphene/Ni(111) interface were performed in different experimental stations in identical experimental conditions allowing for the reproducible sample quality in different experiments. In all experiments the W(110) single crystal was used as a substrate. Prior to preparation of the graphene/Ni(111) system the well-established cleaning procedure of the tungsten substrate was applied: several cycles of oxygen treatment with subsequent flashes to 2300$^\circ$\,C. A well-ordered Ni(111) surface was prepared by thermal deposition of Ni films with a thickness of more than 200\,\AA\ on to a clean W(110) substrate and subsequent annealing at 300$^\circ$\,C. An ordered graphene overlayer was prepared via thermal decomposition of propene (C$_3$H$_6$) according to the recipe described elsewhere~\cite{Dedkov:2008a,Dedkov:2008b,Nagashima:1994,Dedkov:2001,Dedkov:2008}. The quality, homogeneity, and cleanliness of the prepared graphene/Ni(111) system was verified by means of LEED, STM, and core-level as well as valence-band photoemission.

\begin{figure}
\center
\includegraphics[scale=0.4]{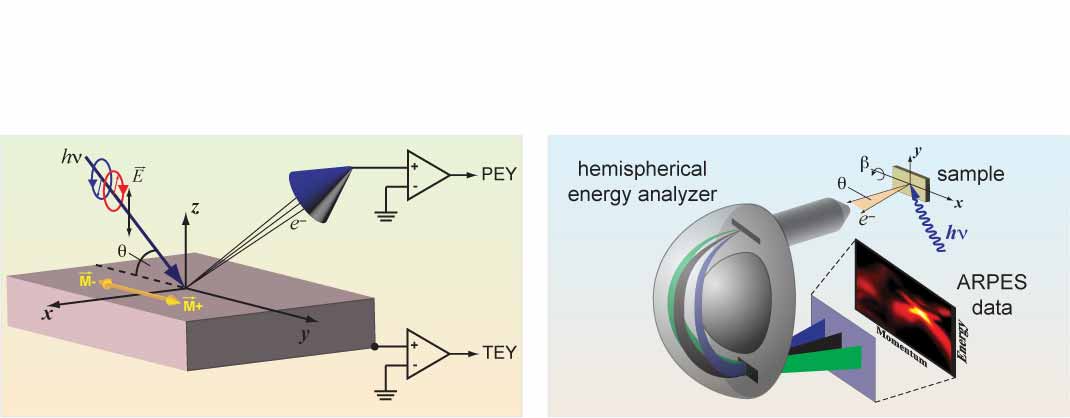}
\caption{Schematic representation of the experimental spectroscopic methods: x-ray absorption spectroscopy and magnetic circular dichroism (left) and angle-resolved photoelectron spectroscopy (right). In XAS or XMCD experiments the photon energy of incoming linearly or circularly polarized light, respectively, is tuned over a particular absorption edge and total or partial electron yield is measured, respectively. XMCD spectra are collected in the remanent magnetic state of the sample. In ARPES measurements the light of a fixed photon energy ($h\nu$) and polarization is used and electrons are analyzed by their kinetic energy, $E_{kin}$, and emission angle, $\theta$, and detected by a 2D CCD detector allowing to measure the sample electronic structure along $k_x$ direction in one shot. Rotation of the sample by angle $\beta$ produces 3D data sets of experimental photoemission intensity, $I(E_{kin},k_x,k_y)$, where $k_y$ is the second in-plane component of the wave-vector calculated from the experimental geometry.}
\end{figure}

STM experiments were carried out in an ultra-high vacuum (UHV) system (base pressure $8\times10^{-11}$\,mbar) equipped with an Omicron variable temperature scanning tunneling microscope. All STM measurements were performed in the constant-current-mode at room temperature using electrochemically etched polycrystalline tungsten tips cleaned in UHV by flash-annealing. The sign of the bias voltage corresponds to the  voltage at the sample. Tunneling parameters are given separately for each STM image: U$_T$ for tunneling voltage and I$_T$ for tunneling current.

XAS and XMCD spectra were collected at the D1011 beamline of the MAX-lab Synchrotron Facility (Lund, Sweden) at both Ni $L_{2,3}$ and C $K$ absorption edges in partial (repulsive potential $U=-100$\,V) and total electron yield modes (PEY and TEY, respectively) with an energy resolution of 80\,meV. Left-hand panel of Fig.\,3 shows the schematic representation of the experimental geometry. Magnetic dichroism spectra were obtained with circularly polarized light (degree of polarization is $P=0.75$) at different angles $\theta$ in the remanence magnetic state of the graphene/Ni(111) system after applying of an external magnetic field of 500\,Oe along the $<1\bar{1}0>$ easy magnetization axis of the Ni(111) film. All absorption measurements were performed at 300\,K. The base pressure during the measurements did not exceed $1\times10^{-10}$\,mbar.

ARPES experiments were performed at the UE56/2-PGM-1 beam-line at BESSY (Berlin, Germany). The experimental station consists of two chambers: preparation and analysis. The sample preparation procedure (oxygen-treatments and flashing of W(110) as well as the preparation of the graphene/Ni(111) system) was performed in the preparation chamber after which sample was transferred into the analysis chamber for further photoemission measurements. The photoemission intensity data sets $I(E,k_x,k_y)$ were collected with a PHOIBOS\,100 energy analyzer (SPECS) while the graphene/Ni(111)/W(110) sample was placed on a 6-axes manipulator (3 translation and 3 rotation axes) (Fig.\,3, right-handel panel). The temperature of the sample during the measurements was kept at 80\,K or 300\,K. The energy/angular resolution was set to 80\,meV/0.2$^\circ$. In case of the spin-resolved PES experiments the mini-Mott-spin-detector (SPECS) was used instead of the 2D CCD detector. The spin-resolved spectra were collected in the remanent magnetic state of the graphene/Ni(111) system (see above) in normal emission geometry. The effective Sherman function was estimated to be $S_{eff}=0.1$ and instrumental asymmetry was accounted via measuring of spin-resolved spectra for two opposite directions of the sample magnetization. The base pressure during all photoemission measurements was below $7\times10^{-11}$\,mbar.

\section{Results and discussion}\label{results}

\subsection{Growth and surface structure of graphene on Ni(111)}

\begin{figure}
\hspace{3cm}
\includegraphics[scale=0.65]{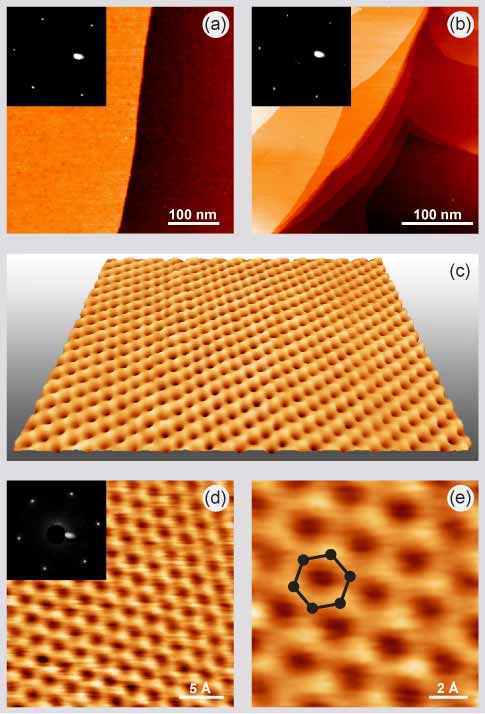}
\caption{(a) Large scale STM image of the W(110) surface showing atomically flat terraces. Tunneling parameters: U$_T$\,=\,0.8\,V; I$_T$\,=\,11\,nA. Inset: a LEED image of the W(110) surface taken at a primary electron energy of 69\,eV. (b) Large scale STM image of the epitaxial Ni(111) layer grown on the W(110) substrate. Tunneling parameters: U$_T$\,=\,0.5\,V; I$_T$\,=\,0.7\,nA. Inset: a LEED image of the Ni(111) surface taken at a primary electron energy of 67\,eV. (c-e) High-quality graphene/Ni(111) system. (c) Large scale constant current STM image of the graphene/Ni(111) surface. Tunneling parameters: U\,=\,0.002\,V; I\,=\,48\,nA. (d) Magnified STM image of the perfect graphene lattice. The inset shows a LEED image obtained at 63\,eV. (e) High magnification STM image showing atomic structure of the graphene monolayer. Tunneling parameters: U$_T$\,=\,0.002\,V; I\,=\,48\,nA). Graphene hexagonal unit cell is marked in (e).}
\end{figure}

\begin{figure}
\center
\includegraphics[scale=0.75]{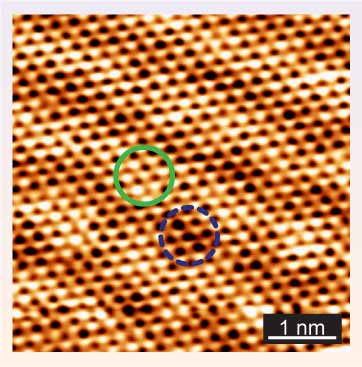}
\caption{Graphene lattice slightly rotated relative to the Ni(111) substrate, showing a moir\'e structure. Areas of different apparent hight are marked in the image by circles. Tunneling parameters: U$_T$\,=\,1.1\,V; I$_T$\,=\,0.18\,nA.}
\end{figure}

In order to check the quality of the samples directly after each preparation step as well as to provide structural details of the graphene sheets at the atomic level, we performed \textit{in situ} STM measurements at room temperature. Figure\,4(a) shows an STM image of the clean atomically flat W(110) surface. The ($1\times1$) LEED pattern of two-fold symmetry [inset in Fig.\,4(a)], typical for the $bcc$ W(110) surface, confirms its high quality. The surface morphology of a 200\,\AA-thick Ni(111) film deposited on W(110) is shown in Fig.\,4(b). Atomically flat terraces separated by steps are visible in the STM image give clear evidence of the epitaxial growth of Ni(111). After the Ni deposition a well-ordered hexagonal ($1\times1$) LEED pattern was observed [inset of Fig.\,4(b)]. Since Ni(111) and graphene have the nearly similar lattice parameters (the lattice mismatch is of only 1.3\,\%), graphene forms the hexagonal ($1\times1$) structure. Fig.\,4(c) shows an overview of a graphene domain on Ni(111) after thermal decomposition of propene. The graphene layer is continuous and exhibits a highly ordered crystallographic structure without any visible defects even over large areas. Fig.\,4(d) represents a magnified topographic image of the graphene lattice together with a typical LEED pattern of monolayer graphene on Ni(111) [inset of Fig.\,4(d)]. A higher magnification STM image of the graphene surface is shown in Fig.\,4(e) with the graphene hexagonal unit cell marked in the image.

So far several possible atomic configurations were considered for the graphene/Ni(111) intrface. Three ``high-symmetry'' structures are known as $hcp-fcc$, $top-hcp$, and $top-fcc$ and they are shown in Figs.\,1(a-c), respectively. In the $top-hcp$ configuration, the C atoms are placed directly above the Ni atoms of the first layer ($top$ site) and the second layer ($hcp$ site). In $top-fcc$, the C atoms are arranged above the Ni atoms of the first and third ($fcc$) layers. In the $hcp-fcc$ configuration, the C atoms are placed above $hcp$ and $fcc$ sites. Three additional configurations were considered recently, which were called $bridge-top$, $bridge-fcc$, and $bridge-hcp$. In these structures, the C atoms are not placed in $hcp-fcc$, $top-hcp$, and $top-fcc$ sites but inbetween~\cite{FuentesCabrera:2008}.

At the moment no clear consensus exists about which of the above described structures is more energetically stable and which kind of structures are observed in experiments. From the theoretical side, G. Bertoni \textit{et al.}~\cite{Bertoni:2005} used density functional theory (DFT) with the Perdew, Burke, and Ernzerhof generalized gradient approximation (GGA-PBE) which yielded the $top-fcc$ as the most stable atomic configuration at the graphene/Ni(111) interface. DFT-PBE studies were also performed by G. Kalibaeva \textit{et al.}~\cite{Kalibaeva:2006} reporting that $top-fcc$ structure is the lowest energy configuration, whereas $hcp-fcc$ is shown to be unstable. The calculations including three additional ``low symmetry'' configurations showed that within DTF with GGA-PBE, none of the structures is stable at the experimentally relevant temperatures; with local-density approximation (LDA), the $bridge-top$ configuration was found to be the most energetically favorable one~\cite{FuentesCabrera:2008}. From the experimental side, R. Rosei \textit{et al.}~\cite{Rosei:1983} and C. Klink \textit{et al.}~\cite{Klink:1995} found that the most stable structure is $hcp-fcc$, whereas Y. Gamo \textit{et al.}~\cite{Gamo:1997} found $top-fcc$ to be the most favorable configuration.

In our case, graphene terraces have a peak-to-peak roughness of 0.2\,\AA\ and show a honeycomb structure with a lattice constant of $2.4\pm0.1$\,\AA\ [Fig.\,4 (d)] which agrees well with the expected 2.46\,\AA\ lattice spacing of graphene. STM images show that in the honeycomb unit cell carbon atoms corresponding to different sites appear with a different contrast, which can be attributed to the differences in the local stacking of the graphene sheet and the Ni(111) substrate. Therefore we interpret our STM images in the following way: Fig. 4 shows a single layer graphene, where carbon atoms most possibly occupy positions corresponding to one of the two non-equivalent configurations -- $top-fcc$ or $top-hcp$. However, it turns to be impossible to directly identify which of the sites are occupied.

Additionally, some different orientations of the graphene relative to the Ni(111) substrate could be observed. Fig.\,5 shows a moir\'e structure, indicating a slight rotation of the graphene layer relative to the Ni(111) substrate demonstrating the simultaneous existence of different stackings in the graphene/Ni(111) system. Two regions showing different apparent heights can be distinguished on the surface (see Fig.\,5). This observation shows that although the interaction between nickel and graphene is relatively strong, different adsorption geometries are locally possible. We would like to note, that such areas represent an very rare case compared with the normal graphene structure as supported by STM, LEED and photoemission measurements.

\subsection{Bonding and magnetism at the graphene/Ni(111) interface}

In order to address the average spatial orientation of selected molecular orbitals (for example $\pi$ or $\sigma$) at the graphene/Ni interface, we vary the sample orientation with respect to the wave vector of the linearly polarized x-ray light and monitor the absorption intensity. The observed changes of the XAS lineshape at the C $K$ edge in the graphene/Ni(111) system represent a nice example demonstrating the so-called \textit{search-light}-like effect~\cite{Stohr:1999b}, which in general can be used for probing of the quadrupole moment of the local charge around the absorbing atom. In such an experiment, the absorption intensity associated with a specific molecular orbital final state has a maximum if the electric field vector is aligned parallel to the direction of maximum charge or hole density, i.\,e. along a molecular orbital, and the intensity vanishes if the electric field vector is perpendicular to the orbital axis. The detailed description of the angular dependence of XAS intensities can be found elsewhere~\cite{Stohr:1999b,Stohr:1999a}.

\begin{figure}
\center
\includegraphics[scale=0.65]{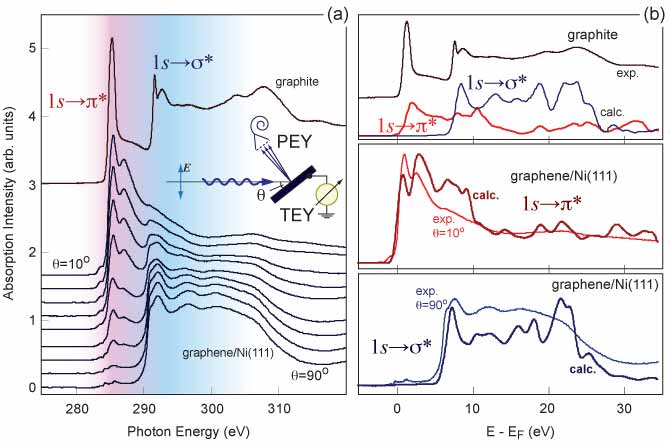}
\caption{(a) Polarization dependence of the absorption at the C $K$ edge of the graphene/Ni(111) system measured as a function of angle, $\theta$ (see Fig.\,3, left panel), between polarization vector of incoming linearly polarized light and the surface normal of the sample~\cite{Weser:2010}. Spectra were collected in the partial electron yield mode and angle was changed with a step of $10^\circ$ from top to bottom. The reference spectrum of pure graphite single crystal is shown in the upper part of the panel for comparison. (b) Comparison between experimental XAS spectra and calculated electron-energy-loss spectra of graphite and graphene/Ni(111) for two different incident angles, $\theta$, where transitions from C $1s$ core level on mostly $\pi^*$- or $\sigma^*$-states occurred. The theoretically calculated spectra are extracted with permission from G. Bertoni, L. Calmels, A. Altibelli, and V. Serin, Phys. Rev. B \textbf{71}, 075402 (2004). Copyright (2004) by the American Physical Society.}
\end{figure}

Figure~\,6\,(a) shows XAS spectra of the graphene/Ni(111) system recorded at the C $K$ absorption edge as a function of the angle, $\theta$, between the direction of the incident linearly polarized x-ray light and the sample surface, e.\,g. between the electrical field vector of light and the sample surface normal (see the inset of Fig.\,6 for the exact geometry of the experiment). The reference XAS spectrum of the graphite single crystal measured at $\theta=30^\circ$ is shown in the upper part of the figure. The spectral features in the two broad regions $283-289$\,eV and $289-295$\,eV can be ascribed to C $1s\rightarrow\pi^*$ and C $1s\rightarrow\sigma^*$ transitions of core electrons into unoccupied states ($\pi^*$, $\sigma^*_1$, and $\sigma^*_2$), respectively. The XAS line shape in both regions is influenced by considerable excitonic effects -- poor core-hole screening~\cite{Bruhwiler:1995,Ahuja:1996,Wessely:2005}. Upon the comparison of the XAS C $1s\rightarrow\pi^*,\sigma^*$ spectrum of the graphene/Ni(111) system with the reference graphite spectrum, considerable changes in the spectral shapes are observed, which can be attributed to a strong chemisorption. A broadening of the $\pi^*$ and $\sigma^*$ resonances gives an evidence for a strong orbital hybridization and electron sharing at the graphene/Ni interface, indicating a strong delocalization of the corresponding core-excited state. A comparison of the present XAS results for graphene on Ni(111) with those recently obtained for graphene/Rh and graphene/Ru~\cite{Preobrajenski:2008} indicates the existing of a very strong covalent interfacial bonding between carbon atoms in the graphene layer and Ni atoms of the substrate.

Both the atomic and the electronic structure of the graphene/Ni(111) system has recently been calculated by G. Bertoni \textit{et al.}~\cite{Bertoni:2005}. The calculation yielded the \textit{top-fcc} configuration of carbon atoms (see Fig.~\,1 and previous section for details) to be energetically the most favorable one and gave a clear indication of the strong interaction between graphene layer and substrate. Similar results were obtained for graphene/Co(0001)~\cite{Eom:2009}. This interaction manifests itself by a considerable modification of the graphene- and Ni-related valence band states as a result of the hybridization of the graphene $\pi$ and the Ni $3d$ states accompanied by a partial charge transfer of spin-polarized electrons from ferromagnetic substrate to graphene. These calculations also predict that several occupied and unoccupied interface states ($I_1\,-\,I_5$) are formed in this system which lead to noticeable modifications of the carbon $K$-edge electron energy loss spectroscopy (EELS) spectrum. The detailed description of the electronic structure of the graphene/Ni(111) system including the interface states will be given in the next section in conjunction with the discussion of the angle-resolved photoemission results.

In the following we would like to compare our XAS results with the recently calculated C $K$-edge EELS spectra for the graphene/Ni(111) interface~\cite{Bertoni:2005}. In this case, experimental XAS spectra taken at $\theta=10^\circ$ and $\theta=90^\circ$ correspond to the calculated EELS spectra for the scattering vector $\mathbf{q}$ perpendicular and parallel to the graphene layer, respectively. The calculated EELS spectra are found to agree well with the experimental XAS data [see Fig.~\,6\,(b)]: (i) the spectra show the same angle (scattering vector) dependence and (ii) the experimentally observed XAS features are well reproduced in the calculated EELS spectra. For example, two peaks in the XAS spectra in the $1s\rightarrow\pi^*$ spectral region at 285.5\,eV and 287.1\,eV of photon energy can be assigned to the double-peak structure in the calculated EELS spectrum at 0.8\,eV and 3.0\,eV above the Fermi level~\cite{Bertoni:2005}. According to the theoretical calculations~\cite{Bertoni:2005}, the first sharp feature in the XAS spectrum is due to the transition of the electron from the $1s$ core level into the interface state $I_4$ above the Fermi level (around the $K$ point in the hexagonal Brillouin zone) which originates from the C\,$p_z$--Ni\,$3d$ hybridization and corresponds to the antibonding orbital between a carbon atom C-$top$ and an interface Ni atom. The second peak in the XAS spectrum is due to the dipole transition of an electron from the $1s$ core level into the interface state $I_5$ above the Fermi level (around the $M$-point in the hexagonal Brillouin zone) which originates from C\,$p_z$--Ni\,$p_x,p_y,3d$ hybridization and corresponds to a bonding orbital between C-$top$ and C-$fcc$ atoms, involving a Ni interface atom. The small feature at 283.7\,eV and the low-energy shoulder in the $1s\rightarrow\pi^*$ absorption spectra probably partly originate from the interface state $I_4$ which is located very close to the Fermi level. In case of the XAS C $1s\rightarrow\sigma^*$, the theory also correctly describes the shape of the absorption spectra~\cite{Bertoni:2005}.

\begin{figure}
\center
\includegraphics[scale=0.5]{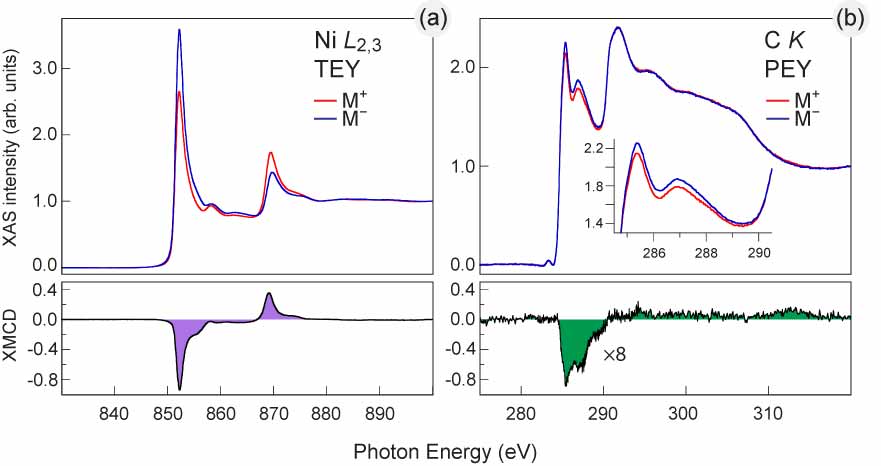}
\caption{XMCD spectra of the graphene/Ni(111) system~\cite{Weser:2010}: absorption spectra measured with circularly polarized light for two opposite orientations of the sample magnetization are shown in the upper part for the Ni $L_{2,3}$- (a) and C $K$-edges (b). The corresponding differences reflecting the strength of the dichroic signal are shown in the lower part of the respective figures.}
\end{figure}

The strong hybridization between graphene $\pi$ and Ni $3d$ valence band states at the graphene/Ni(111) interface leads to the partial charge transfer of the spin-polarized electrons from Ni onto C with the appearance of an induced effective magnetic moment of carbon atoms~\cite{Bertoni:2005} which can be detected in an experiment which is sensitive to the magnetic state of particular element, like XMCD. Figure\,7 shows XAS spectra of the graphene/Ni(111) system obtained for two opposite magnetization directions with respect to the polarization of the incident X-ray beam (upper panels) together with the resulting XMCD signal (lower panels). The spectra collected at the Ni $L_{2,3}$ edge in TEY mode and at the C $K$ edge in PEY mode are presented in the left-hand panel and in the right-hand panel, respectively. The Ni $L_{2,3}$ XMCD spectrum (white line as well as fine structure behind the absorption edge) is in  perfect agreement with previously published spectroscopic data~\cite{Srivastava:1998,Dhesi:1999,Nesvizhskii:2000}. The quantitative analysis of the absorption spectra obtained on a magnetic sample with circularly polarized light can be performed with the help of the so-called sum rules for spin- and orbital-magnetic moments~\cite{Thole:1992,Carra:1993}. The intensities of $L_3$ and $L_2$ absorption lines and their differences for the parallel and anti-parallel orientations of the projection of photon spin on the sample magnetization direction are quantitatively related by the sum rules to the number of $3d$ holes in the valence band of the ferromagnetic material and the size of the spin and orbital magnetic moments. 

From the experimental Ni $L_{2,3}$ TEY XAS data (Fig.\,7) the bulk values of Ni magnetic moments were derived, using a number of $3d$ holes in the valence band of Ni $n_h\,=\,1.45$~\cite{Sorg:2004} and polarization of light $P\,=\,0.75$. At room temperature and in TEY mode (more bulk sensitive), Ni provides a spin moment of $\mu_S=0.69\,\mu_B$ and an orbital moment of $\mu_L=0.07\,\mu_B$, respectively. These values coincide with the previously published experimental results~\cite{Baberschke:1996,Srivastava:1998}. The experimentally obtained spin-magnetic moment is very close to the calculated bulk value of $\mu_S=0.67\,\mu_B$ for the graphene/Ni(111) system~\cite{Bertoni:2005}. For the most energetically favorable configuration of carbon atoms on Ni(111), $top-fcc$, the calculations predict a reduction of the spin-magnetic moments of Ni interface atoms by 16\,\% down to $0.56\,\mu_B$~\cite{Bertoni:2005}. The experimental data collected at the Ni $L_{2,3}$ absorption edge in the PEY mode (more surface/interface sensitive) also shows a slight reduction of the spin moment to $\mu_S=0.63\,\mu_B$. However, the observed decreasing is not so pronounced as yielded by the theoretical calculation, which can be explained by the large contribution of the bulk-derived signal to the XMCD spectra.

The most important and interesting results of these XMCD experiments on the graphene/Ni(111) system is the observation of the relatively large dichroic contrast at the C $K$ absorption edge [Fig.\,7(b)]. In order to magnify the measured magnetic contrast at the $1s\rightarrow\pi^*$ absorption edge, these XMCD spectra were collected in the PEY mode with the circularly polarized light at an angle of $\theta=20^\circ$. We note that the observed differences in the XAS spectra collected at this angle visible in Figs.\,6 and 7 are due to the different polarization of light: in Fig.\,6(a) the data are obtained with the linearly polarized x-rays, i.\,e. the strong angular dependence of the absorption signal is due to the different graphene's orbital orientation; whereas the data in Fig.\,7 are taken with the circularly polarized light, i.\,e. both $1s\rightarrow\pi^*$ and $1s\rightarrow\sigma^*$ transitions are nearly equivalently possible. The C $K$ XMCD spectrum reveals that the major magnetic response stems from transitions of the $1s$ electron onto the $\pi^*$-states, while transitions onto the $\sigma^*$-states yield very weak (if any) magnetic signal. These results indicate that only the C $2p_z$ orbitals of the graphene layer are magnetically polarized due to the hybridization with the Ni $3d$ band. The sharp structure at the $1s\rightarrow\pi^*$ absorption edge originates from hybridized C\,$p_z$--Ni\,$3d$ and C\,$p_z$--Ni\,$p_x,p_y\,3d$ states (see earlier discussion and Ref.~\cite{Bertoni:2005}).

The appearance of XMCD signal at the C $K$-edge shows that indeed the Ni film induces a magnetic moment in the graphene overlayer. However, at the C $K$ absorption edge, the electron transitions occur from the non-spin-orbit split $1s$ initial states to the $2p$ final states and thus, in the analysis of the dichroism effect at the $K$ edge one equation in the selection rules is missed. This means that the XMCD signal at $K$ edges provides the information only on the orbital magnetic moment $\mu_{orb}$~\cite{Thole:1992,Carra:1993,Huang:2002}. The partial charge transfer from Ni to C atoms in the graphene/Ni(111) system was calculated for the 22-atom\,(graphene)/Ni cluster~\cite{Yamamoto:1992} yielding 0.205$e^-$ per C atom in graphene which leads to the $2p$-orbital occupation number of $n_p\,=\,2.205 e^-$. Using the the C $K$ XAS spectra the procedure described in work~\cite{Huang:2002}, the orbital magnetic moment of $\mu_{orb}=1.8\pm0.6\times10^{-3}\,\mu_B$ per C atom was extracted. The relatively large uncertainty arises mainly  from the estimation of the number of C $2p$ holes, the background subtraction of XAS spectra, and from the error for the degree of circular polarization of light. 

The theoretical work~\cite{Bertoni:2005} also gives the values for the spin magnetic moment of $-0.01\,\mu_B$ and $0.02\,\mu_B$ for C-$top$ and C-$fcc$ atoms, respectively. However, the magnetic splitting of the majority and minority parts of the interface states $I_3$ and $I_4$ was found between 0.13 and 0.55\,eV, respectively, which may yield higher values for the magnetic moment. Due to the impossibility to directly extract the value of the spin magnetic moment form the $K$ edge XMCD spectra, we apply a simple comparison with the magnetic measurements on similar systems in oder to estimate the average $\mu_S$ value for the carbon atoms at the graphene/Ni(111) interface. For the Fe/C multilayers clear magnetic signals of carbon were obtained by using the resonant magnetic reflectivity technique~\cite{Mertins:2004}. Hysteresis loop recorded at C $K$ absorption edge gave a clear proof of ferromagnetism of carbon atoms at room temperature with a measured spin magnetic moment of $\mu_S=0.05\,\mu_B$ induced by adjacent Fe atoms. The observed ferromagnetism of carbon in the Fe/C multilayered system was related to the hybridization between the Fe $3d$ orbitals and the C $p_z$ orbitals which are normal to the graphene-type layered $sp^2$-coordination. The second comparison can be performed with carbon nanotubes on ferromagnetic Co substrate~\cite{Cespedes:2004}. Carbon nanotubes were shown to become magnetized when they are in contact with magnetic material. Spin-polarized charge transfer at the interface between a flat ferromagnetic metal substrate and a multiwalled carbon nanotube leads to a spin transfer of about $0.1\,\mu_B$ per contact carbon atom. Additionally, a comparison of the XMCD spectra obtained at the C $K$ edge in graphene/Ni(111) (present work) and at the O $K$ edge in O/Ni(100)~\cite{Sorg:2006}, reveals the approximately same magnitude of the XMCD signal. For the O/Ni(100) system, where the induced spin-magnetic moment of $0.053\,\mu_B$ on oxygen atom was calculated, the theoretically simulated XAS and XMCD spectra agree well with the experimental data. Considering these analogous systems, we estimate the induced magnetic moment for graphene on Ni(111) to be in the range of 0.05-0.1 $\mu_B$ per carbon atom.

The experimentally observed effective magnetic moment of carbon atoms of the graphene layer on Ni(111) is also confirmed by our spin-resolved photoemission results. Figure\,8 shows the spin-resolved valence-band spectra (a) and the corresponding spin polarization as a function of the binding energy (b) of pure Ni(111) and the graphene/ Ni(111) system. These spectra were recorded with the photon energy of $h\nu$\,=\,65\,eV at room temperature in normal emission geometry. The spin-resolved PES spectra of the pure thick bulk-like Ni(111) film on W(110) is in very good agreement with previously reported data presenting the clear spin-contrast in the valence band region and a spin polarization value of about $-\,60\%$ at the Fermi level. The presence of graphene on Ni(111) strongly modifies the valence band spectrum of Ni indicating the strong interaction between valence band states of graphene and Ni (a thorough discussion of valence band photoemission spectra will be given in the next section.). In the graphene/Ni(111) system the spin polarization of Ni $3d$ states at $E_F$ is strongly reduced to about $-\,25\%$. The considerable modifications of the spin-resolved structure of Ni $3d$ states as well as the reduction of the spin polarization at $E_F$ could be considered as an indication of the decreasing of magnetic moment of Ni atoms at the graphene/Ni(111) interface. In general, the Ni $3d$ photoemission signal around $E_F$ consists of the sum of the bulk- and surface-derived photoemission intensities. But due to the fact that the presented spectra were collected in the most surface-sensitive regime (kinetic energy of electrons is $E_{kin}\approx60$\,eV and the corresponding inelastic mean free path is $\lambda\sim5$\,\AA) the main contribution in these spectra comes from the surface Ni atoms.  The same behavior for the surface Ni layer was predicted by theory~\cite{Bertoni:2005} and experimentally observed in the presented XMCD measurements.  

\begin{figure}
\center
\includegraphics[scale=0.6]{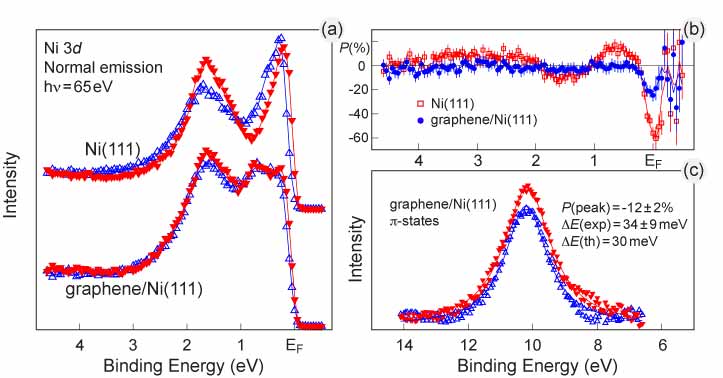}
\caption{(a) Spin-resolved PES spectra and (b) corresponding spin polarization of Ni $3d$ valence band states as a function of the binding energy for Ni(111) and the graphene/ Ni(111) system. (c) Spin-resolved photoemission spectra of the $\pi$-states of graphene on Ni(111). A spin polarization value of about $(-12\pm2\%)$ together with a considerable $\pi$-band exchange splitting of about 34\,meV are observed. All spectra were collected in the normal emission geometry with the photon energy of 65\,eV.}
\end{figure}

Fig.\,8(c) shows the spin-resolved spectra of the $\pi$ states of graphene on Ni(111) measured in normal emission geometry with the photon energy of 65\,eV. These states are located at 10.1\,eV of binding energy (see next section for details). After the careful subtraction of the background spin-polarization which originates from the spin-resolved Ni $3d$ valence band states the clear spin contrast could be detected for the graphene $\pi$ states with the maximum spin-polarization of $P=-12\pm2\,\%$ at room temperature. The two spin-resolved components are fitted with Lorenzian-shape profiles giving the exchange splitting of about $34\pm9$\,meV for these state which agrees well with the value extracted from theoretical work~\cite{Bertoni:2005}. A comparison of the spin-resolved spectra with the background spin polarization originating from Ni $3d$ states shows that the spin moment of graphene is aligned antiparallel to the magnetization of Ni meaning antiparallel magnetic coupling of graphene to Ni. However, this conclusion is based only upon the normal emission spin-resolved measurements and has to be clarified by more detailed spin-resolved PES experiments. Here we would like to note, that the experimental evidence of a pronounced spin polarization at room temperature together with the splitting of the $\pi$ states in the graphene/Ni(111) system are in contradiction to the results previously reported by O. Rader \textit{et al.}~\cite{Rader:2009}, where the absence or very small spin-polarization of the graphene $\pi$ states was observed. The origin of this discrepancy is however not clear at the moment and further spin-polarized PES investigations of the graphene/Ni(111) system should be undertaken in order to resolve this issue.

\subsection{Electronic properties of graphene on Ni(111)}

\begin{figure}
\center
\includegraphics[scale=0.55]{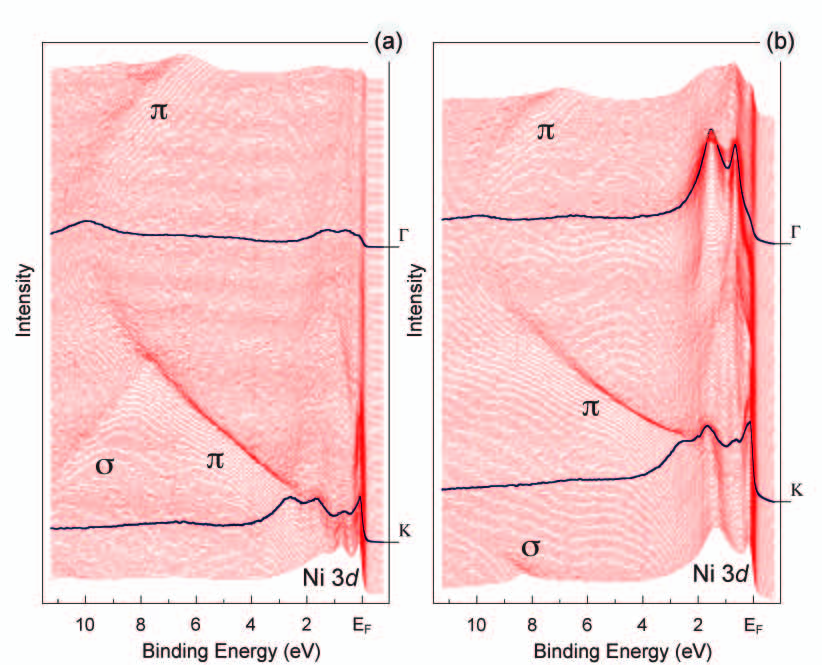}
\caption{Angle-resolved photoemission spectra of the graphene/Ni(111) system recorded along the $\Gamma-K$ direction of the hexagonal Brillouin zone at (a) 70\,eV and (b) 100\,eV of photon energy. Spectra corresponding to $\Gamma$ and $K$ points are marked by thick black lines.}
\end{figure}

Fig.\,9 shows two series of angle-resolved photoemission spectra measured along the $\Gamma-K$ direction of the hexagonal Brillouin zone of the graphene/Ni(111) system. Each series is extracted from the 3D sets of data of photoemission intensity $I(E_B,k_x,k_y)$, where $E_B$ is the binding energy and $k_x,k_y$ are the orthogonal components of the in-plane wave vector. For the graphene/Ni(111) system the $K$ and $M$ points of the Brillouin zone are reached at $1.7$\,\AA$^{-1}$ and at $1.4$\,\AA$^{-1}$, respectively. Photoemission data presented in Fig.\,9 were collected at two different photon energies: 70\,eV (a) and 100\,eV (b). The variation of the photon energy, i.\,e. photoemission cross-section, gives a possibility to clearly separate the partial contributions of emissions from graphene-derived and Ni-derived valence band states in the spectra. The presented photoemission data are in very good agreement with previously published results~\cite{Nagashima:1994,Shikin:2000,Dedkov:2001,Dedkov:2008a,Gruneis:2008}, but acquired with a much better energy/wave-vector resolution allowing for a more accurate analysis of photoemission dispersions. In Fig.\,9 one can clearly discriminate dispersions of graphene $\pi$- and $\sigma$-derived states in the region below $2$\,eV of the binding energy (BE) as well as Ni $3d$-derived states near $E_F$. The binding energy of the graphene $\pi$ states in the center of the Brillouin zone (in the $\Gamma$ point) equals to $10.1$\,eV which is approximately by $2.4$\,eV larger than the binding energy of these states in pure graphite. The shift to larger binding energy is different for $\sigma$ and $\pi$ valence band graphene-derived states. This behavior can be explained by the different hybridization strength between these states and Ni $3d$ valence band states which is larger for the out-of-plane oriented $\pi$ states compared with the one for the in-plane oriented $\sigma$ states of the graphene layer. The binding energy difference of $\approx2.4$\,eV for the $\pi$ states and $\approx1$\,eV for the $\sigma$ states between graphite and graphene on Ni(111) is in good agreement with previously reported experimental and theoretical values~\cite{Dedkov:2008a,Bertoni:2005}. The effect of hybridization between Ni $3d$ and graphene $\pi$ states can be clearly demonstarted in the region around the $K$ point of the Brillouin zone: (i) one of the Ni $3d$ bands at 1.50\,eV changes its binding energy by $\approx150$\,meV to larger BE when approaching the $K$-point; (ii) a hybridization shoulder is visible in photoemission spectra which disperses from approximately 1.6\,eV to the binding energy of the graphene $\pi$ states at the $K$ point (see also Fig.\,10 for a detailed view). The strong hybridization observed in PES spectra underlines the fact that the $\pi$ states might become spin-polarized and might gain a non zero-magnetic moment due to charge transfer from the Ni atoms to the carbon atoms. 

\begin{figure}
\center
\includegraphics[scale=0.5]{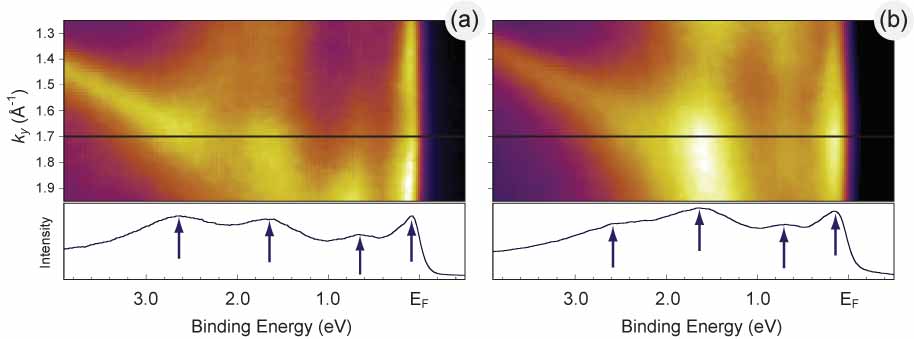}
\caption{Regions around the $K$ point of photoemission intensity of the graphene/Ni(111) system measured at (a) 70\,eV and (b) 100\,eV of photon energy. The corresponding intensity profiles at the $K$ point are shown in the lower panels with arrows indicating the main photoemission features discussed in the text.}
\end{figure}

Considering the electronic band structure of the graphene/Ni(111), the region around the $K$ point delivers the most interesting and important information with respect to the possible spin-filtering effects in the graphene/ferromagnet or ferromagnet/graphene/ferromagnet sandwich-like structures. This part of the electronic structure measured with two different photon energies ($h\nu=70$\,eV and $100$\,eV) is shown in Fig.\,10 as color maps (upper panels) together the corresponding intensity profiles directly at the $K$ point (lower panels). Firstly, the spectral function of the graphene layer on Ni(111) is characterized by the absence of well-ordered structure of the graphene $\pi$-bands in the vicinity of the Fermi level and secondly, the Dirac-cone is not preserved. Both observations can be attributed to a strong interaction between graphene layer and metallic substrate leading to a strong hybridization between the graphene $\pi$ and the Ni $3d$ valence band states. In the vicinity of the $K$ point a number of photoemission peaks can be clearly distinguished: (i) a sharp peak about the Fermi level at $0.1-0.2$\,eV BE, (ii) a graphene $\pi$-states-related peak at $2.65$\,eV BE, (iii) two peaks at $0.7$\,eV and $1.65$\,eV BE.

In the following we perform a detailed analysis of the experimentally obtained electronic structure relying mainly on two comprehensive sets of electronic structure calculations currently available for the graphene/Ni(111) system~\cite{Karpan:2008,Bertoni:2005}. The calculations by G. Bertoni \textit{et al.}~\cite{Bertoni:2005} predict the existence of three interface states below the Fermi level originating from the strong hybridization between the Ni $3d$ and the graphene $\pi$ states and corresponding to: ($I_1$) bonding between C-$fcc$ and interface Ni atoms; ($I_2$) bonding between C-$top$ and interface Ni atoms; ($I_3$) antibonding between C-$fcc$ and interface Ni atoms. V. M. Karpan \textit{et al.}~\cite{Karpan:2008} performed the band structure calculations of the graphene/Ni(111) system with the major emphasis on the investigation of the spin-dependent transport properties of the Ni/graphene/Ni sandwiches. Both calculations yield a quite complicated band structure of the graphene/Ni(111) system around the Fermi level due to the strong hybridization between the graphene and the Ni valence band states. From the analysis of the region around the $K$ point of the hexagonal Brillouin zone we could distinguish a number of flat bands which are clearly separated from each other. The positions of the bands taken from both calculations~\cite{Karpan:2008,Bertoni:2005} are summarized in Table\,1 with the assignment of the particular band character.

\begin{table}
\caption{Binding energies (in eV) of the main valence band features around the $K$ point extracted from the calculated band structures of the graphene/Ni(111) system. Positions of experimental photoemission peaks are listed in the right-hand column.}
\begin{indented} 
\item[]\begin{tabular}{@{}ccc|ccc|c}
\br
\multicolumn{3}{c|}{Karpan \textit{et al.}~\cite{Karpan:2008}}&\multicolumn{3}{c|}{Bertoni \textit{et al.}~\cite{Bertoni:2005}}&{Experiment}\\
\mr
spin $\uparrow$&spin $\downarrow$& &spin $\uparrow$&spin $\downarrow$& &\\
\mr
0.7&0&Ni\,$3d$&0.8&0&Ni\,$3d$&0.1-0.2\\
1.2&0.9&Ni\,$3d$&1.5&0.9&Ni\,$3d$&0.7\\
0.1&-0.2&Ni\,$3d$ -- gr.\,$\pi$&2.8&2.1&Ni\,$3d$&1.65\\
 &1.8&Ni\,$3d$ -- gr.\,$\pi$&0.2&-0.18&$I_3$&2.65\\
 & & &2.4&1.96&$I_2$& \\
 & & &3.37&3.24&$I_1$& \\ 
\br
\end{tabular}
\end{indented} 
\end{table}

The interpretation of the experimentally observed photoemission features around the Fermi level could be performed as presented in the following. The photoemission peak close to the Fermi level (0.1-0.2\,eV BE) could be considered as a combination of the interface state $I_3$ (both spins) with a large contribution of the graphene $\pi$-character and the Ni\,$3d$($\downarrow$)-band. The second peak at $0.7$\,eV BE could be assigned to the combination of the Ni\,$3d$($\uparrow$)- and Ni\,$3d$($\downarrow$)-bands present in both calculations (first and second rows in Table\,1, respectively). The feature at $1.65$\,eV could be considered as a combination of Ni\,$3d$($\uparrow$)-band and $I_2$($\downarrow$)-state with a large graphene $\pi$-character. The last photoemission peak (2.65\,eV BE) could be assigned to the interface state $I_2$($\uparrow$) with large contribution of the graphene $\pi$-character.

In order to check the theoretical predictions concerning the CPP spin-dependent electronic transport properties of the ideal graphene/Ni(111) interface, we perform a careful analysis of the constant energy photoemission maps close to the Fermi level. We would like to admit that such an analysis can be rather complicated due to the fact that Ni $3d$ bands, which dominate the photoemission intensity around the Fermi level, are very flat in the vicinity of $E_F$. Figure\,11 shows the constant energy cuts of the 3D data set obtained at $h\nu$=100\,eV for the graphene/Ni(111) system. These energy cuts were taken at (a) $4$\,eV and (b) $2.7$\,eV of BE as well as at (c) the Fermi level. The energy cut at $E_B=4$\,eV shows characteristic photoemission intensity patterns of the graphene layer which reflect the symmetry of the system. Below the Dirac  point (crossing of straight dispersion lines of $\pi$ states in free-standing graphene) the graphene $\pi$ bands are visible in the first Brillouin zone whereas no bands can be seen in the second one. Additionally several energy bands are present in the middle of the Brillouin zone (outlined by dashed yellow lines) which also show hexagonal symmetry. These bands originate from the hybridization of the Ni and graphene valence band states. The constant energy cut taken in the region of the minimal binding energy of the graphene $\pi$ states ($E_B=2.7$\,eV) is shown in Fig.\,11(b). In the case of graphene/Ni(111) the Dirac point is not preserved due to the strong hybridization of Ni $3d$ and graphene $\pi$ states around the $K$ point. This can also be directly recognized at this energy cut where graphene $\pi$ states produce broad intensity spots instead of sharp points in the wave-vector space. As in the previous case, we observe a number of valence band states in the middle part of the Brillouin zone which again could be assigned to the hybridization-derived states.

\begin{figure}
\center
\includegraphics[scale=0.475]{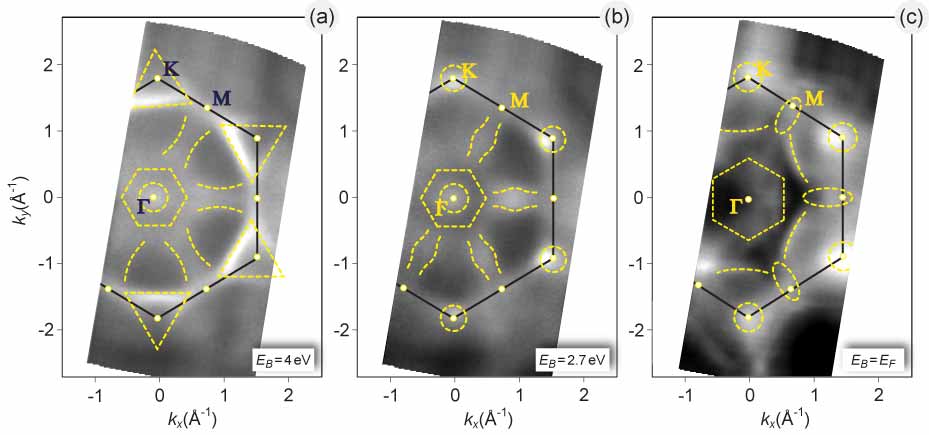}
\caption{Constant energy cuts of the 3D data sets in the energy-wave vector space, $I(E_B,k_x,k_y)$, obtained via a $\beta$-scan of the graphene/Ni(111) system obtained at 100\,eV of photon energy. The energy cuts are taken at (a) $4$\,eV and (b) $2.7$\,eV BE as well as at (c) $E_F$. As a supplementary we also present a movie which chows the binding energy scan through the valence band of the graphene/Ni(111) system with corresponding energy cuts in the wave-vector space. Dashed lines are guides to the eye.}
\end{figure}

The most interesting and important information in view on the spin-dependent transport properties of the graphene/Ni(111) system, can be extracted form the constant energy cut obtained at the Fermi energy which is presented at Fig.\,11(c). Already the analysis of Figs.\,9(a,b) and 10(a,b) shows that the photoemission intensity is increased around the $K$ point and along the $K-M$ direction of the hexagonal Brillouin zone, that correlates with the increased photoemission intensity observed in the energy cut shown in Fig.\,11(c) for the Fermi energy. Additionally, a number of arcs surrounding the $K$ points and weak (but distinguished) diamond-shape regions of increased intensity is clearly visible in the middle part and around the $M$ points of the Brillouin zone, respectively. Upon the comparison of the obtained photoemission results for the graphene/Ni(111) system (Figs.\,9--11) with the band structure calculations for this system (Fig.\,2)~\cite{Karpan:2007,Karpan:2008,Bertoni:2005}, we find very good agreement between theory and experiment. Particularly, the region around the Fermi level for the ideal graphene/Ni(111) system is well reproduced in the experiment, confirming the main predictions of the theory. Unfortunately, at the present stage of the experiment, we can not specify the spin-character of energy bands, which should be the subject of future spin- and angle-resolved photoemission investigations.

\section{Conclusions}

In conclusion, structural, magnetic and electronic properites of the high-quality graphene/ferromagnet interface [graphene on Ni(111)] were investigated by means of scanning tunneling microscopy, x-ray absorption spectroscopy and magnetic circular dichroism as well as via mapping of the band structure by means of angle-resolved photoemission. STM investigation shows that perfectly ordered epitaxial graphene layers can be prepared by elevated temperature decomposition of hydrocarbons, with domains larger than the terraces of the underlying Ni(111). A strong modification of the electronic structure of the graphene layer and Ni substrate upon graphene adsorption on the ferromagnetic substrate was detected by all spectroscopic methods. This modification is due to the considerable hybridization of the graphen $\pi$ and Ni $3d$ valence band states accompanied by the partial charge transfer of spin-polarized electrons from Ni onto C atoms leading to the appearance of the effective magnetic moment in the graphene layer. The presence of an effective magnetic moment on carbon atoms of the graphene layer was unambigously proven by XMCD and spin-resolved photoemission measurements. The experimentally obtained electronic structure of occupied and unoccupied states was compared with available band structure calculations allowing the clear assignment of spectral features in the XAS and ARPES data. The good agreement between theory and experiment was also found upon the analysis of the Fermi energy cuts, that give us an opportunity to confirm the main statements of the theoretical works predicting the spin-filtering effects of the graphene/Ni(111) interface. However, the clear assignment of the spectroscopic valence band features have to be performed in future spin-resolved photoemission experiments.

\section*{Acknowledgements}

We would like to thank all co-workers and collaborators for their contributions to this work, in particular, K. Horn, M. Weser, S. B\"ottcher, C. Enderlein, A. Preobrajenski, E. Voloshina, M. Sicot, P. Leicht, A. Zusan. We are also grateful to P. Kelly and L. Calmels for fruitful discussions. This work has been supported by the European Science Foundation (ESF) under the EUROCORES Programme EuroGRAPHENE (Project ``SpinGraph''). M. F. gratefully acknowledges the financial support by the Research Center ``UltraQuantum'' (Excellence Initiative), by the Deutsche Forschungsgemeinschaft (DFG) via the Collaborative Research Center (SFB) 767, and the Baden-W\"urttemberg Stiftung.

\section*{References}


\end{document}